\documentclass[a4paper,12pt]{article}
\usepackage{graphics}
\usepackage{graphicx}
\usepackage{amsmath}

\title
{Inter-arrival time distribution for the non-homogeneous Poisson process}
\author{Gleb Yakovlev\footnotemark[1], John B. Rundle\footnotemark[1], Robert Shcherbakov\footnotemark[1],\\
and Donald L. Turcotte\footnotemark[2]}

\begin{document}
\footnotetext[1]{Center for Computational Science and Engineering, University of California, One Shields Ave. Davis, CA 95616}
\maketitle
\footnotetext[2]{Department of Geology, University of California, One Shields Ave. Davis, CA 95616}
\maketitle
\begin{abstract} We derive an analytical expression of the inter-arrival time distribution for a non-homogeneous Poisson process (NHPP). This expression is exact and is applicable to any time interval, finite or infinite. As an illustration, we present simulation results for three different intensity functions.

\end{abstract}

The Poisson process has found numerous applications in science, engineering, economics and other areas. The NHPP is probably the best known generalization of the Poisson process (see for example [1]). It is characterized by a deterministic intensity function that describes how the rate of the process changes in time. For an ordinary Poisson processes, this function is a constant. The NHPP has many important characteristics, but in this work we concentrate on the distribution of inter-arrival times, also known as a waiting time distribution, a somewhat ``cumulative'' property of the process. For an ordinary Poisson process, this distribution is an exponential if the process is observed over a sufficiently large time interval. In the general case of the NHPP, various approximations were developed; examples include the time-dependent Poisson process, when intensity changes gradually with time, and the conditional Poisson process, when the intensity function has a probabilistic interpretation.

Let us state the central question of this work: Given that the NHPP starts at time $0$ and that we observe events during a fixed (finite or infinite) interval $T$, what is the distribution of inter-arrival times between consecutively arriving events?

We denote the arrival times of the NHPP by $t_0=0,\; t_1,\; \dots,\; t_n$. The corresponding inter-arrival times are $s_1=t_1,\; s_2=t_2-t_1,\; \dots,\; s_n=t_n-t_{n-1}$. Let $\lambda(t)$ (intensity function) be a known, deterministic function of time and let $\Lambda(t)=\int_0^t \lambda(u) du < \infty$ be the mean value function (expected number of events of the NHPP on the time interval $(0,t)$). The total expected number of events in the interval $T$ is $\Lambda(T)$. Let $N(t)$ be the number of arrivals on the interval $(0, t)$. The probability of having $n$ arrivals in the interval $(t, t+x)$ is given by [1]
\begin{equation}\label{prob}
    Prob\{N(t+x)-N(t)=n\} = \frac{[\Lambda(t+x)-\Lambda(t)]^n}{n!} e^{-[\Lambda(t+x)-\Lambda(t)]}
\end{equation}
where conventionally we set $0!=1$.

First we write an equation for the probability of the first inter-arrival time $s_{1}>x$ \textit{\textbf{and}} the total number of events $N(T)=n$
\begin{align}
    A(x,n) &= Prob\{s_{1}>x \;\bigcap\; N(T)=n\} \notag \\
    &= \frac{1}{\Lambda(T)} e^{-\Lambda(x)} \times \frac{[\Lambda(T)-\Lambda(x)]^n}{n!} e^{-[\Lambda(T)-\Lambda(x)]}\label{term-A} \\
    &= \frac{[\Lambda(T)-\Lambda(x)]^n}{n!}  \frac{e^{-\Lambda(T)}}{\Lambda(T)} \notag
\end{align}
where $1/\Lambda(T)$ is a normalization constant (inverse of the expected number of events on the interval $T$), $\exp(-\Lambda(x))$ is the probability to have no arrivals on the interval $(0,x)$ and the second factor is the probability to have exactly $n$ arrivals on the interval $(x,T)$. These probabilities should be multiplied, as $N(b-a)$ and $N(d-c)$ are independent random variables on disjoint intervals $(a,b)$ and $(c,d)$ [1].

Next, we write an expression for the probability of the $k^{th}$ arrival time $t_k\in(y,y+dy)$, $k > 0$ \textit{\textbf{and}} the $k^{th}+1$ inter-arrival time $s_{k+1}>x$ \textit{\textbf{and}} the total number of events $N(T)=n$
\begin{align}
    d B(x,y,k,n) &= Prob\{t_k\in(y,y+dy)\; \bigcap\; s_{k+1}>x \;\bigcap \;N(T)=n\}\notag\\
    &= \frac{1}{\Lambda(T)} \frac{[\Lambda(y)]^{k-1}}{(k-1)!} e^{-\Lambda(y)}\lambda(y)dy \notag\\
    &\times e^{-[\Lambda(y+x)-\Lambda(y)]} \label{term-B} \\
    &\times \frac{[\Lambda(T)-\Lambda(y+x)]^{n-k}}{{n-k}!} e^{-[\Lambda(T)-\Lambda(y+x)]} \notag\\
    &= \lambda(y)dy \frac{[\Lambda(y)]^{k-1}}{(k-1)!} \frac{[\Lambda(T)-\Lambda(y+x)]^{n-k}}{{n-k}!}
            \frac{e^{-\Lambda(T)}}{\Lambda(T)}\notag
\end{align}
where the first factor is the probability to have the $k^{th}$ arrival time $t_k\in(y,y+dy)$ [1], the second factor is the probability to have no arrivals on the interval $(y,y+x)$, and the last is the probability to have $n-k$ arrivals on the interval $(y+x,T)$. Again, these probabilities should be multiplied due to mutual independence.

The total probability to have an inter-arrival time $S>x$, $G(x)$, is given by
\begin{equation}\label{G}
    G(x)=Prob\{S>x\} = \sum_{n=1}^\infty A(x,n) + \sum_{n=2}^\infty \sum_{k=1}^{n-1} \int_0^{T-x} d B(x,y,k,n)
\end{equation}
where the upper limit of integration is $T-x$, because for any given $x<T$, $y+x$ must be less then $T$. Interchanging the order of integration and the two summations ($\lambda(t)$ is integrable function), summing up by $k$ and $n$, integrating over $y$ and by parts, we arrive at
\begin{equation}\label{G}
    G(x) = \frac{1}{\Lambda(T)}\int_0^{T-x} \lambda(y+x) e^{-[\Lambda(y+x)-\Lambda(y)]} dy
\end{equation}
We see that this probability is properly normalized, $G(0)=1$ and $G(T)=0$. The probability density function (pdf) for
Equation (\ref{G}) is given by $-G_x^{'}(x)$ and (after integrating by parts) is equal to
\begin{equation}\label{pdf}
    pdf(x) = \frac{1}{\Lambda(T)}\int_0^{T-x} \lambda(y)\lambda(y+x) e^{-[\Lambda(y+x)-\Lambda(y)]} dy + \frac{\lambda(x)
    e^{-\Lambda(x)}}{\Lambda(T)}
\end{equation}
This equation can be interpreted as follows: The factor $\lambda(y+x) exp\{-[\Lambda(y+x)-\Lambda(y)]\}$ is the pdf to have the inter-arrival time $x$ given that the sum of all previous $s_i$ is $y$. $\lambda(y)/\Lambda(T)d t$ is the probability for the NHPP to be found in the interval $(y,y+dy)$, $y>0$. This means that at least one arrival of NHPP has already happened. In other words, the first arrival density is missing in the integral, which is exactly compensated by the second term in Equation (\ref{pdf}).

\begin{figure}\label{pp}
    \includegraphics[]{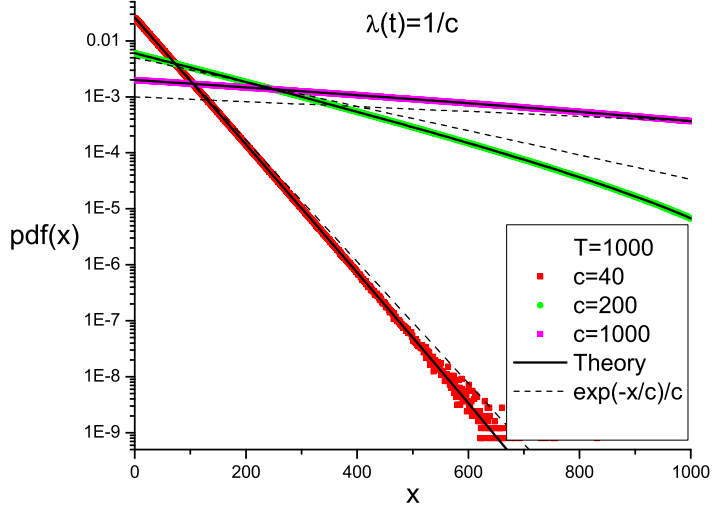}
    \caption{pdf for $\lambda(t)=1/c$. Solid lines are from Equation (\ref{pdf-pp}) and dashed lines are for the corresponding exponential distributions.}
\end{figure}

To illustrate Equation (\ref{pdf}) we perform numerical simulations for three intensity functions $\lambda(t)$: $\lambda(t)=1/c$ for an ordinary Poisson process, $\lambda(t)=c/(1+t)$ for a decaying intensity function, and $\lambda(t)=c t$ for a growing intensity function. These functions were chosen because the integrals in (\ref{pdf}) can be taken explicitly. Our simulations were done as follows: For every intensity function we ran $10^8$ processes. During every process, the next inter-arrival time $s_{i+1}$ was drawn from a known function (via the Inverse Transform method for NHPP, see [1]) which depends on the sum of all previous $s_i$'s. This inter-arrival time was utilized for statistics unless the corresponding arrival time $t_{i+1} > T$, in which case the current process was terminated and the next was started.

On all our figures simulation results are presented in colors while theoretical pdf's from (\ref{pdf}) are given by solid black lines. The distribution of inter-arrival times for $\lambda(t)=1/c$ is given by
\begin{equation}\label{pdf-pp}
    pdf(x)=\frac{T+c-x}{cT}e^{-x/c}
\end{equation}
and examples are shown in Figure (1) for three different values of $c$. This is a homogeneous Poisson process with the mean time between events $c$. Dashed lines represent three corresponding exponential distributions. The deviations from an exponential distribution are easily understood as follows: given that the process is observed at time $y < T$ (arrival time $t_i = y$), the inter-arrivals time $s_{i+1}>T-y$ cannot be observed because the corresponding arrival times will exceed the observational interval $T$. In other words, the finiteness of $T$ introduces memory into the process.
\begin{figure}
    \label{omori}
    \noindent\includegraphics[]{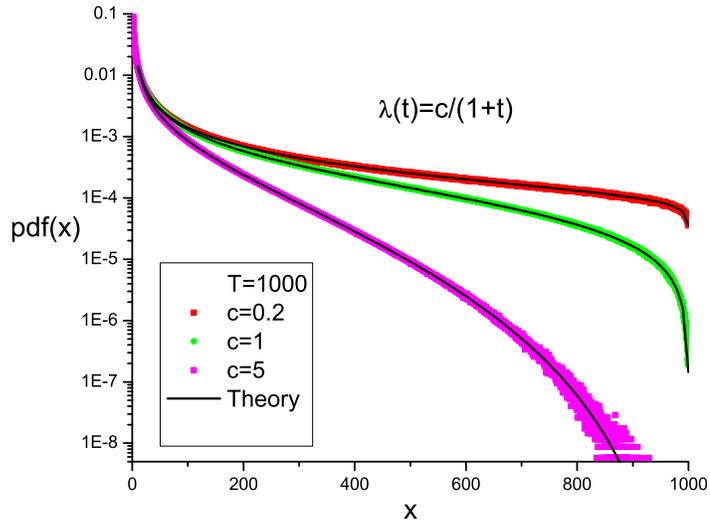}
    \caption{pdf for $\lambda(t)=1/c$. Solid lines are from Equation (\ref{omori-pp}).}
\end{figure}

Our next example is for the intensity function $\lambda(t)=c/(1+t)$. The corresponding inter-arrival pdf is given by
\begin{equation}\label{omori-pp}
    pdf(x)=\frac{(1-\frac{x}{1+T})^c - (1+x)^{-1-c}}{x \ln(1+T)}
\end{equation}
and simulation results are shown in Figure (2).

The last intensity function to be considered here is $\lambda(t)=c t$. Simulation results are presented in Figure (3). An explicit expression for the inter-arrival pdf can be easily found using Mathematica and is not presented here because of it's length.
\begin{figure}
    \label{linear}
    \noindent\includegraphics[]{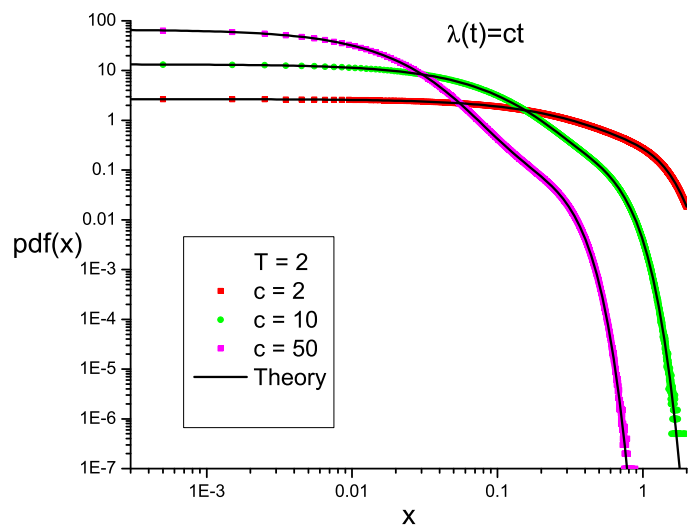}
    \caption{pdf for $\lambda(t)=c t$.}
\end{figure}

In this work we obtained an exact expression for the inter-arrival time distribution for NHPP. Results of this work will be applied to the distribution of earthquake aftershock waiting times [2].

We would like to thank Alexander Burin, Sayan Basu, and Alexander Soshnikov for useful discussions and comments. This work has been supported by DOE Grant DE-FG02-04ER15568 (GM, JBR, RS) and NSF Grant ATM 0327558.

Electronic address: gleb@cse.ucdavis.edu

[1] S.M. Ross, Probability Models, Academic Press, eighth edition, 2003.

[2] R. Shcherbakov, G. Yakovlev, D.L. Turcotte, and J.B. Rundle, A model for the distribution of aftershock waiting times (to be published).
\end{document}